\documentclass[longbibliography,aps,pre,twocolumn,groupedaddress]{revtex4-2}

\usepackage{graphicx}
\usepackage{natbib}
\usepackage{subfigure}
\usepackage{color}
\usepackage{multirow}
\usepackage{amssymb}
\usepackage{booktabs}

\usepackage[hidelinks]{hyperref}
\graphicspath{{figs/}}

\begin{document}

\title{On the origins of reverse Janssen effect}

\author{Srujal Shah}
\email{srujal.shah@lut.fi}
%\homepage[]{Your web page}
%\thanks{}
%\altaffiliation{}
\affiliation{School of Energy Systems, Lappeenranta-Lahti University of Technology LUT, 53851 Lappeenranta, Finland}

\author{Ana Maria Mosquera Gomez}
\email{ana.mosquera.gomez@lut.fi}
%\homepage[]{Your web page}
%\thanks{}
%\altaffiliation{}
\affiliation{School of Energy Systems, Lappeenranta-Lahti University of Technology LUT, 53851 Lappeenranta, Finland}

\author{Payman Jalali}
\email{payman.jalali@lut.fi}
%\homepage[]{Your web page}
%\thanks{}
%\altaffiliation{}
\affiliation{School of Energy Systems, Lappeenranta-Lahti University of Technology LUT, 53851 Lappeenranta, Finland}

\author{Lou Kondic}
\email{kondic@njit.edu}
%\homepage[]{Your web page}
%\thanks{}
%\altaffiliation{}
\affiliation{Department of Mathematical Sciences and Center for Applied Mathematics and Statistics, New Jersey Institute of Technology, Newark, New Jersey 07102, USA}

\date{\today}

\begin{abstract}  
We consider experimentally and computationally the phenomenon of the reverse Janssen effect, involving the counterintuitive finding that the force on the base of a column containing granular particles may be larger than the weight of the granular material itself.  This finding is in contrast to the common Janssen effect, for which the force on the base is smaller than the particle weight, illustrating one of the best-known differences between granular and liquid systems. We find that the reverse Janssen effect is strongly influenced by the pouring protocol: under Earth's gravitational field, we find that the reverse Janssen effect is strongly and consistently influenced by the pouring height, as well as by (to somewhat lesser degree) pouring flux. Pouring grains from the height measured in tens of particle diameters leads to a reverse Janssen effect that is an order of magnitude stronger than the one found for small pouring heights of few particle diameters. This influence of the particles' delivery protocol allows for the development of a better understanding of the general features of the reverse Janssen effect and of the comparison between experiments and simulations reported in this and previous works. 
\end{abstract}

\maketitle

\section{Introduction}
\label{sec:introduction}
 Studies of stresses exerted by granular media are extensive, due to their vital role in numerous applications~\cite{duran2012,liu95,jaeger96,mueth98}. As a classical example, in a vertical column of granular material confined by sidewalls, such as a silo, the base of the column experiences a pressure exerted by grains that do not depend linearly on the filled height of the column~\cite{duran2012,janssen95,vanelprl00}. The distribution of wall stresses was originally studied and formulated by Janssen~\cite{janssen95} for applications in silos, which typically have a diameter and a height significantly larger than the grain size. In Janssen's type of analysis~\cite{duran2012}, computations based on the force balance for a static cylindrical column of granular material of a circular cross-section of radius $R$, at any depth $h$ leads to the expression for the pressure $p=({\rho_{\mathrm{b}} g R/{2K\mu}}) \left[1-\exp^{-2Kh \mu/R}\right]$, where $\rho_{\mathrm{b}}$ is the bulk density of granular medium, $g$ is the gravitational acceleration, and $\mu$ is the coefficient of friction between particles and the wall. Also, $K$ is the coefficient of redirection of vertical stress toward the wall, or in other words, the ratio of horizontal to vertical stresses. The expression above leads to a well-known saturation of the pressure at the column base, with the remaining weight of the material supported by the side walls.  Such saturation is one of the most obvious differences between granular and liquid systems. 

In the original Janssen's analysis~\cite{janssen95} (see also~\cite{tighe07}), the ratio of the column size to particle diameter is infinitely large. In recent decades, however, numerous studies have used experimental setups involving smaller columns relevant to lab-scale devices. These studies have often found significant discrepancies to Janssen's formula~\cite{bertho03,bratberg05}. In addition to the scale of the system, the filling protocol also influences the Janssen effect. For instance, three types of experimental results were reported in Ref.~\cite{vanel99} corresponding to two filling protocols and a tapping protocol, with different results for each. In another study of the Janssen effect in a setup involving moving side wall~\cite{perge12}, the variation of apparent mass versus filled mass was investigated to demonstrate the features of the Janssen effect while particles were discharging, rather than being statically packed in a column. 
Another study reported in Ref.~\cite{ovarlez05} shows that the wall force is indeed strongly influenced by the wall's motion, with the wall forces even larger than the weight of the deposited material.
The reported results~\cite{vanel99,bertho03,perge12,ovarlez05} show that the type of rearrangement for grains driven by tapping or moving of the wall is key in developing the particle-wall forces; complementary results were also found by discrete element method (DEM) simulations~\cite{landry04,windowsyule19}.
 
 Recent work~\cite{ciamarra_prl20} has shown that the pressure at the column base could be larger than the material weight/area.  This 'reverse' Janssen effect, obtained in narrow columns and for small filling heights (in terms of particle diameter) is counterintuitive, since it leads to the base pressure which is even higher than what one would expect for a corresponding amount of liquid placed in the column.  In~\cite{ciamarra_prl20}, the reverse Janssen effect was identified both in experiments and in DEM simulations, although experimental results have shown a significantly stronger reverse Janssen effect compared to simulations. To our knowledge, the source of these significant differences between the experiment and simulations was left unclear.  

 In this paper, we consider, via experiments and DEM simulations, the origins of the reverse Janssen effect.  We focus in particular on the (small) filling heights such that the reverse Janssen effect is relevant, and do not discuss in any detail larger filling heights for which the usual Janssen effect is dominant.  As we will show, a combination of careful and well-documented experiments and simulations, including precisely defined experimental and simulation protocols, allows us to clarify the source of the reverse Janssen effect itself, and also to outline the possible sources of significant differences between experiments and simulations, reported although not discussed in Ref.~\cite{ciamarra_prl20}, and also observed in our work. 

 \section{Methods}
 \label{sec:methods}

 \subsection{Experimental Methods} 
 
\begin{figure}[!ht]
\centering \resizebox{0.95\hsize}{!}{\includegraphics{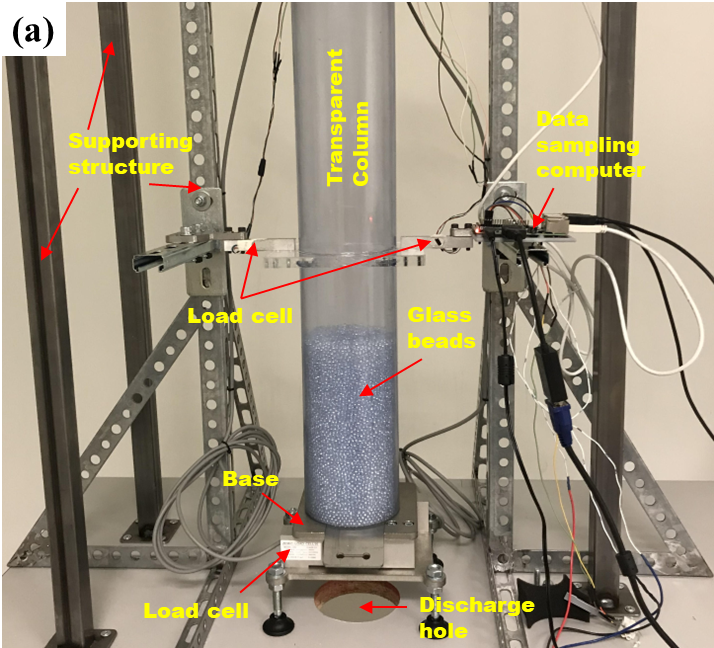}}
\centering \resizebox{1\hsize}{!}{\includegraphics{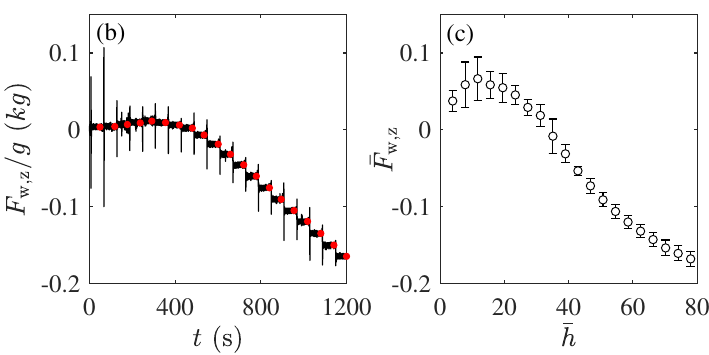}}
\caption{(a) Schematic view of the experimental setup.
  (b) Instantaneous values of measured wall force (in the vertical, z, direction) versus time during a single experiment of step-wise filling of the $D = 52$ mm tube using $d = 4$ mm glass beads. Red circles represent the mean values computed after force oscillations decay. (c) Evolution of mean wall force (circles) versus the height of loaded material averaged over 5 realizations of the experiment from part (b);  error bars represent the standard deviation. In part (c) and the subsequent figures, the results are presented in nondimensional form as described in the text.}
\label{fig:experiment}
\end{figure}

 Figure~\ref{fig:experiment}a provides a schematic representation of the experimental setup. A polymethyl methacrylate (PMMA) cylindrical tube of length 60 cm is positioned vertically. The tube is mounted above the base, maintaining a gap of about 0.5 mm between the lower end of the tube and the base. Hence, the force exerted on the wall and the bottom can be measured independently. For this purpose, two strain gauges, measuring the total vertical wall force on the tube,  are attached to the tube symmetrically. In addition, two strain gauges are symmetrically attached to the bottom plate to measure the force transmitted from the granular column to the base; the sampling frequency for all gauges is 2000 Hz. The reliability of the measured mass-force (force divided by gravity) is $0.001$ kg and $0.0023$ kg for the sensors located at the wall and basal positions, respectively.  

 The column is filled through consecutive steps by pouring (using a conical hopper) a specified amount of granular material into the tube of diameter $D$, followed by a waiting time of $60$ s before the next loading. The pouring is performed from the top of the tube (height of 60 cm measured from the base of the tube)  in each filling step with a flux of approximately $0.01$ kg/s. A mean value of the wall force is found at each filling step by averaging the measured force within the last 20 seconds of the step. To ensure the filling height of granular material ($h$) after each loading step remains independent of $D$, the added mass per step is scaled proportionally to $D^2$, with the step-filling masses  $m_s = 0.036,~0.050$ and $0.223$ kg for the three considered tube diameters of $44$, $52$, and $105$ mm. The consecutive filling continues until the tube is filled by a mass $M_\mathrm{L}(D)$. Note that in the following sections, we present the results in nondimensional form, with nondimensional quantities indicated by overbars for clarity. We use particle diameter, $d$, as the length scale and the weight of the loaded mass as the force scale. These are the only two scales required for our purposes.

 % exp_setup.jpg

The particles used in this work are commercially obtained spherical glass beads with mean diameters of $d = 2$, $3$, and $4$ mm. We conducted experiments using a tube with a diameter of $D = 52$ mm for all three particle diameters, considering only monodisperse samples. Additionally, we performed experiments with $d = 4$ mm particles in two other tube diameters, $D = 44$ mm and $D = 105$ mm. For the tube of $D = 52$ mm with $d = 4$ mm particles, experiments with roughened walls were also included.

\subsection{Simulation Methods} 

We use LIGGGHTS~\cite{Kloss2012}, which solves the translation and rotation equations of motion for individual particles with the linkage to the interparticle contact model for resolving the contact forces. The normal forces are given by Hertzian interaction potential (3/2 power for three-dimensional simulations considered here), and frictional effects are modeled using the Cundall-Strack model~\cite{cundall79}. The details about the contact models can be found in the literature~\cite{tsuji92,direnzo04}. The material parameters are governed by the experiments; see Table~\ref{tbl:table_sim} for details; an earlier work~\cite {Shah2020} used the same parameters. The filling of the column follows a protocol that resembles the experimental one as well; spherical particles representing glass beads fall under gravity once they are released from the top of the column at a random location, with the particle flux matching the experimental value given before. After each filling step, particles are allowed to relax for a sufficient time so that the magnitude of their velocities decays to essentially zero before starting the next filling step.  

\begin{table}[!ht]
\small
  \caption{Simulation parameters corresponding to the physical experiments.  These parameters, with a particle diameter of 4 mm, define the `reference case' in the simulations; these parameters are used if not specified otherwise.  The friction coefficients correspond to the case with smooth walls in experiments. The subscripts used are as follows: wall - w, particle - p, base - b.}
  \label{tbl:table_sim}
  \begin{tabular*}{0.48\textwidth}{@{\extracolsep{\fill}}llll}
    \hline
    Mechanical & \hspace{-0.2 cm} Value & \hspace {0.2 cm} Mechanical &  \hspace{-0.5 cm} Value  \\
    properties &  & \hspace {0.2 cm} properties  &  \\
    \hline
    Young's modulus, GPa &  & \hspace {0.2 cm} Friction coefficient & \\
    \hspace {0.1 cm} base & \hspace{-0.2 cm} $200$ & \hspace {0.3 cm} $\mu_{\rm p,b}$ & \hspace{-0.5 cm} $0.45$\\
    \hspace {0.1 cm} particle & \hspace{-0.2 cm} $10$ & \hspace {0.3 cm} $\mu_{\rm p,p}$ & \hspace{-0.5 cm} $0.4$\\
    \hspace {0.1 cm} wall & \hspace{-0.2 cm} $3$  & \hspace {0.3 cm} $\mu_{\rm p,w}$ & \hspace{-0.5 cm} $0.3$\\\\
    Poisson's ratio &  & \hspace {0.2 cm} Restitution coefficient & \\
    \hspace {0.1 cm} base & \hspace{-0.2 cm} $0.3$ & \hspace {0.3 cm} $e_{\rm p,b}$ & \hspace{-0.5 cm} $0.8$ \\
    \hspace {0.1 cm} particle & \hspace{-0.2 cm} $0.2$ & \hspace {0.3 cm} $e_{\rm p,p}$ & \hspace{-0.5 cm} $0.9$ \\
    \hspace {0.1 cm} wall & \hspace{-0.2 cm} $0.4$ & \hspace {0.3 cm} $e_{\rm p,w}$ & \hspace{-0.5 cm} $0.85$ \\\\
    \hline
    Simulation & \hspace{-0.2 cm} Value & \hspace {0.2 cm} Simulation &  \hspace{-0.5 cm} Value  \\
    parameters &  & \hspace {0.2 cm} parameters  &  \\
    \hline
    Density, kg/m$^3$ &  & \hspace {0.2 cm} Diameter, mm &  \\
    \hspace {0.1 cm} particle & \hspace{-0.2 cm} $2500$ & \hspace{0.3 cm} particle  &  \hspace{-0.5 cm} $2$, $3$, $4$ \\
    Height of column, cm & \hspace{-0.2 cm} $30$ & \hspace {0.2 cm} DEM time step, s  & \hspace{-0.5 cm} $5\times10^{-7}$ \\
    \hline
  \end{tabular*}
\end{table}

\section{Results}
\label{sec:results}

We start by briefly presenting experimental results and then follow by a detailed discussion of the results of simulations; later in Sec.~\ref{sec:discussion}, we discuss the sources of the differences in the results between simulations and experiments. In the presentation of the results, a particular configuration is chosen as the reference case, and various parameters are varied one at a time. The parameters for the reference case are assumed unless specified otherwise. The reference case is specified by the particle diameter $d$ of 4 mm, tube diameter $D$ of 52 mm, particle physical properties as specified in Table~\ref{tbl:table_sim}, smooth domain boundaries in experiments, and particles released from the top of the domain (height of the tube).

\begin{figure}[thb]
\centering \resizebox{1\hsize}{!}{\includegraphics{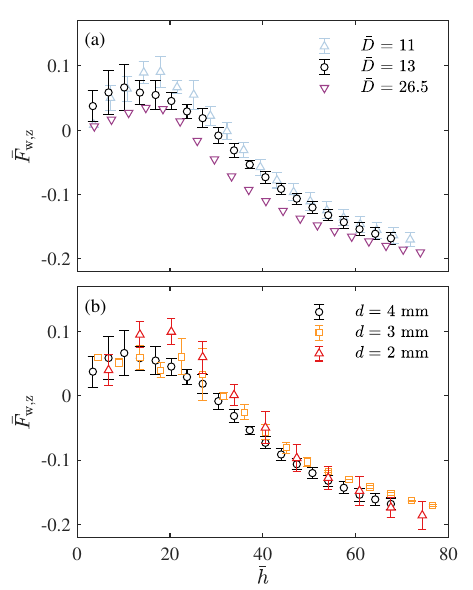}}
\caption{Experimental wall force as a function of $\bar h$; (a) the particle diameter $d = 4$ mm is fixed, and the tube diameter $D$ is varied, and (b) $D$ is fixed at  52 mm and $d$ is varied. Average values over three realizations are reported. Error bars represent the standard deviation, except for $\bar{D} = 26.5$ (purple triangle in panel (a)), where the variation is smaller than the symbol size.}
\label{fig:exp_force_varD_var_dp2}
\end{figure}

Figure~\ref{fig:experiment}(b, c) shows the results for the side-wall force, $\bar{F}_{\rm w,z}$ as a function of time (b) and filling height (c). Note that the positive values of $\bar{F}_{\rm w,z}$ correspond to the reverse Janssen effect (the net wall force acting up, in the $+z$ direction, where ${\rm z}$ is the vertical coordinate with the origin at the base). Here, the magnitude of the reverse Janssen effect is about 10\%, which is consistent with the experimental results presented in Ref.~\cite{ciamarra_prl20}.   Figure~\ref{fig:exp_force_varD_var_dp2} shows how the results depend on (a) $D$ as $d$ is kept fixed, and (b) $d$ as $D$ is kept fixed. While the trend of the results in Fig.~\ref{fig:exp_force_varD_var_dp2}(a) is as expected (wider tubes lead to weaker reverse Janssen effect), the results shown in Fig.~\ref{fig:exp_force_varD_var_dp2}(b) are more elaborate and require further discussion, which is more informative when presented jointly with the discussion of simulation results, presented later in the paper.

\begin{figure}[thb]
\centering \resizebox{1\hsize}{!}{\includegraphics{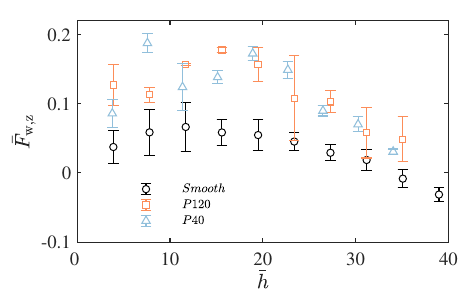}}
\caption{Experimental wall forces as a function of filling height for different wall roughnesses of the tube of $D = 52$ mm. Note the different range of the axes compared to Fig.~\ref{fig:exp_force_varD_var_dp2}; the experiments with roughened walls were carried out for $\bar h < 40$. 
}
\label{fig:exp_force_varFrictionp}
\end{figure}

Figure~\ref{fig:exp_force_varFrictionp} shows the wall force for experiments in which tube wall roughness is increased. A smooth wall is used for the reference case, here referred as "smooth". The other cases are identified by the standard code of grit sandpapers used to roughen the inner surface of the tube walls, which are P120 and P40. The former has a mean particle diameter of abrasive particles about 125 microns, whereas the latter corresponds to 425 microns. We have not attempted to explicitly measure the coefficient of friction but note that, as the grit number decreases, the surface roughness increases, and therefore, the particle-wall friction coefficient is expected to increase. We find that,  within the error bars, roughening with either of these sandpapers affects the peak of wall force equally, increasing it by about a factor of two when compared to the smooth tube. Therefore, the reverse Janssen effect is boosted by increasing the friction between particles and the wall. These results will also be discussed in the context of simulation results later in the paper.

\begin{figure}[!th]
\centering \resizebox{1\hsize}{!}{\includegraphics{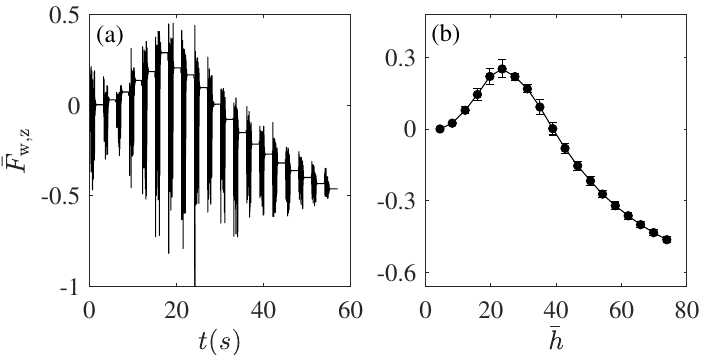}}
\caption{(a) Simulated values, averaged over 1/2000 s, of wall force in the ${\rm z}$ direction exerted by the particles as a function of time. (b) The wall force in the ${\rm z}$ direction, as a function of the height of loaded material for the same simulation values as in part (a),  averaged over three realizations with error bars corresponding to standard deviations. In this and the following figures, the computed data are shown by symbols, and the line segments between the symbols are shown for visualization purposes only.
} 
\label{fig:simulations}
\end{figure}

Figure~\ref{fig:simulations} depicts the simulation results obtained using the protocol described in Sec.~\ref{sec:methods}. We note that all the material parameters are nominally the same as in the experiments, and also the particles are released in the simulations consistently with the experimental protocol (albeit for practical reasons from smaller pouring height, 30 cm in simulations versus 60 cm in the experiment); see Movie 1 in the Supplemental Material ~\cite{sup_mat_mov} demonstrating the filling protocol for the reference case showing the velocity magnitude of the particles. While the results exhibit a similar pattern between the experiments and simulations, the numerical values of the force, particularly the peak values of the reverse Janssen effect, are noticeably {\it larger} in simulations compared to the experimental results.  These findings are inconsistent with the simulation results in~\cite{ciamarra_prl20}, where the peak of the reverse Janssen effect in simulations is significantly {\it smaller} than in the experiments.  As discussed in Sec.~\ref{sec:height}, this inconsistency will aid in developing a better understanding of the reverse Janssen effect. In the subsequent discussion,  we will also point out the possible sources of the difference between the simulations and the experiments, both in the present work and in Ref.~\cite{ciamarra_prl20}.

\begin{figure}[!th]
\centering \resizebox{1\hsize}{!}{\includegraphics{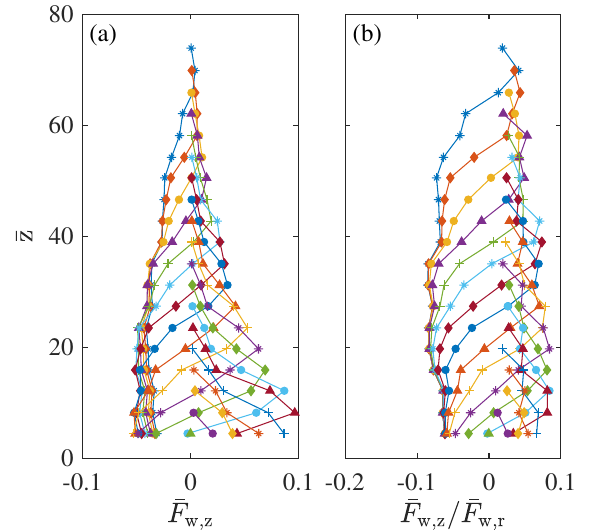}}
\caption{(a) Wall force, $\bar F_{\rm {w,z}}$, exerted by the particles in the ${\rm z}$ direction as a function of distance to the base, $\bar {\rm z}$. (b) Wall force normalized by the radial force. The lines' color is used for visualization purposes only.
}
\label{fig:wall_force}
\end{figure}

Regarding the origin of the reverse Janssen effect,  simulations are particularly useful since they allow for easy extraction of the forces along the side wall as a function of the distance from the base.  Figure~\ref{fig:wall_force}(a) shows the wall forces in the z direction after each filling. After the first filling, the wall forces exerted by the particles in the z direction are summed up, obtaining the total force plotted in the figure at the z value corresponding to the top of the first layer.  After the second filling, we separately sum up the wall forces from each layer (note that the forces of the first layer particles are now different due to the presence of the second layer),  and plot the forces for each layer as two symbols at the z coordinates corresponding to the top of the first and second layers.  This process is continued for all the consequent layers,  allowing to visualize the evolution of wall forces in the z direction as the column is filled. Figure~\ref{fig:wall_force}(b) shows the wall forces in the z direction normalized by the wall forces in the radial direction. 
%\end{document}

Figure~\ref{fig:wall_force} shows that for small filling heights, and close to the free surface, the ${\rm z}$-component of the wall force is positive (leading to the reverse Janssen effect).  We also observe that 
for larger values of ${\rm z}$, the positive peak of $\bar F_{\rm{w,z}}$ decreases in magnitude. This decrease is the key to understanding the influence of the pouring height which has a dominant effect on the magnitude of the reverse Janssen effect, as discussed in the next section.

\subsection{Influence of pouring height}
\label{sec:height}

The simulation results presented so far were obtained by releasing the particles from the fixed height (measuring from the base).  Therefore, as the tube is filled with particles, the pouring height (measured from the top of the granular surface) decreases as the pouring process proceeds. Figure~\ref{fig:wall_force}(a) shows that the reverse Janssen effect also decreases as the pouring height decreases.

\begin{figure}[!ht]
\centering \resizebox{1\hsize}{!}{\includegraphics{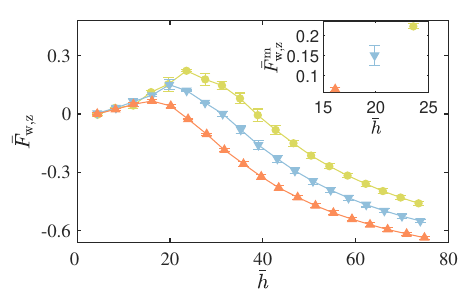}}
\caption{ Wall force for different pouring heights, $H$: 16.5 cm (yellow), 8.5 cm (blue), and 3.5 cm (orange).  Here, $H$ is measured from the surface of the granular matter already poured (so, relative to the base, the height at which the particles are released increases as the filling proceeds). The inset shows how the maximum values of wall force $\bar{F}_{\rm w}^m$ increase as $H$ is increased.
}
\label{fig:force_height}
\end{figure}

To illustrate the pouring height effects, the pouring protocol was modified to release the particles consistently from the same height, $\bar H$ above the granular surface. Figure~\ref{fig:force_height} shows the wall forces for three values of $\bar H$.  It is evident that a reduction in $\bar H$ significantly decreases the magnitude of the reverse Janssen effect.

The influence of pouring height on the magnitude of the reverse Janssen effect provides deeper insight into its origins.  Essentially, the larger $\bar H$ is, the larger the kinetic energy of the falling particles.  Upon impact, this energy leads to the motion of the already deposited particles in the outward direction; these particles interact with the side walls and produce net upward force, which is the reverse Janssen effect. The produced wall force is directed upwards due to broken-up symmetry at the side walls: the pressure due to gravitational compaction is higher deeper in the sample, and thus, the particles that are pushed out towards the wall are biased to move on average upward. This upward motion leads to a net upward force.  

While the described source of the reverse Janssen effect is basically similar to the one outlined in~\cite{ciamarra_prl20}, we note that the reverse Janssen effect that we discuss here is an {\it order of magnitude larger} than the one reported in~\cite{ciamarra_prl20} (the latter one is similar in magnitude to our results when pouring from $H = 3.5$ cm).  {\it Therefore, the kinetic energy of the falling particles, determined predominantly by the pouring height, is the main factor determining the magnitude of the reverse Janssen effect.  }

\begin{figure}[!ht]
\centering \resizebox{1\hsize}{!}{\includegraphics{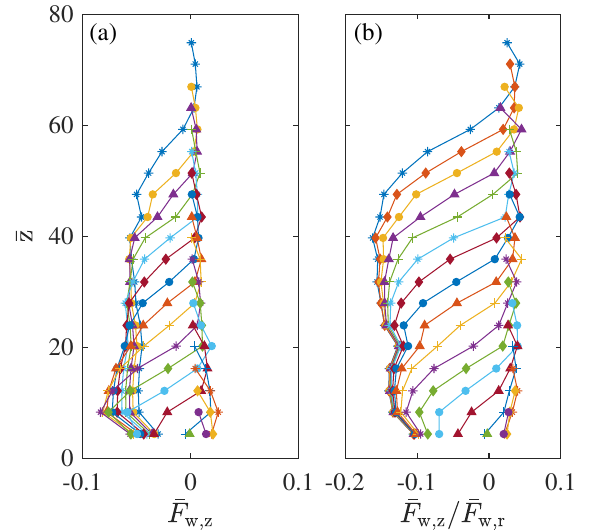}}
\caption{Wall force for small pouring height of $H = 3.5$ cm for each filling (a) Wall force, $\bar F_{\rm {w,z}}$, exerted by the particles in the ${\rm z}$ direction as a function of distance to the base, $\bar {\rm z}$. (b) Wall force normalized by the radial force.
The lines' color is used for visusalization purposes only.
}
\label{fig:wall_force_small_height}
\end{figure}

Figure~\ref{fig:wall_force_small_height} further illustrates the influence of pouring height: here, we plot the wall forces as a function of ${\rm z}$ for $H = 3.5$ cm.  Direct comparison with Fig.~\ref{fig:wall_force} shows a significant difference and in particular, Fig.~\ref{fig:wall_force_small_height} does not show large values of $\bar F_{\rm w,z}$ for small values of $\bar {\rm z}$: this is because $\bar H$ is always small, and therefore the  kinetic energy of the poured particles has only weak effect on the magnitude of the wall forces responsible for the reverse Janssen effect.

We continue by exploring the parameter space with the goal of further clarifying the role of the pouring protocol and various parameters on the features of the reverse Janssen effect.  

\subsection{Parameteric dependence}
\label{sec:parametric}

\subsubsection{Pouring protocol}

In the previous section, the results illustrating the crucial influence of the pouring height were presented.  Here, the discussion continues around important details of the protocol, which gain relevance. This exploration will further clarify some aspects of the reverse Janssen effect.
 
In the simulation results presented so far, the particles were introduced from the top of the domain (for the reference case) by simply letting the particles fall down under gravity.  However, in real physical experiments, the particles may also be given random impulses in the radial direction. In light of our understanding of the source of the reverse Janssen effect when the pouring height is significant, one may wonder whether such an additional component may play a role. For this purpose, a stochastic radial velocity of magnitude of 0.1 $\rm m/s$ to the pouring particles at the time of their release is added.

\begin{figure}[!ht]
\centering \resizebox{1\hsize}{!}{\includegraphics{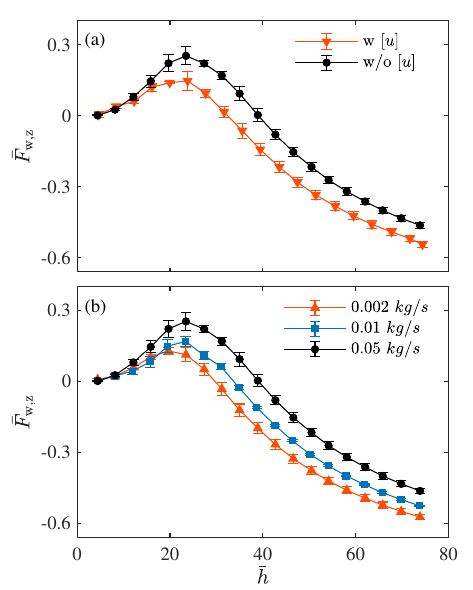}}
\caption{(a) Wall force in the simulations without (black) and with (red) an additional radial component of particle velocity (see also Movie 2 in the Supplemental Material~\cite{sup_mat_mov}). (b) Wall force in the simulations where particle flux is modified. Note that the black filled circles correspond to the reference case in this and the following figures.
}
\label{fig:pouring}
\end{figure}

Figure~\ref{fig:pouring}(a) shows the wall forces for the reference case and the case when an additional stochastic velocity component is added.  Clearly, there is a strong effect, with a much smaller reverse Janssen effect in the latter case.  Our explanation is that this additional component makes particles lose energy due to the inelasticity of the collisions, consequently leading to a weaker influence on the already deposited particles upon deposition.  

Another parameter relevant to the pouring protocol is particle flux. Figure~\ref{fig:pouring}(b) shows the results where the flux is modified by changing the time period during which the particles are released. A significantly stronger reverse Janssen effect can be observed in the reference case, suggesting that the source of the effect is collective: a significant rate of energy transfer is needed to produce positive side wall forces.  

 We have also explored the role of the size of the release region and the amount of particles released during each pouring event. These quantities did not influence the reverse Janssen effect in any appreciable manner (results not shown for brevity). 

\subsubsection{Influence of particle and wall properties }

Regarding particle and wall properties, we limit the discussion to friction, elasticity, and Young's modulus.  We will see that each of these quantities influences the magnitude of the reverse Janssen effect, and understanding their impacts enhances our comprehension of the effect itself.

Friction between the particles and between the particles and the side wall is clearly an important parameter since, in particular, the friction between the particles and the side walls is directly responsible for the reverse Janssen effect.

\begin{figure}[!ht]
\centering \resizebox{1\hsize}{!}{\includegraphics{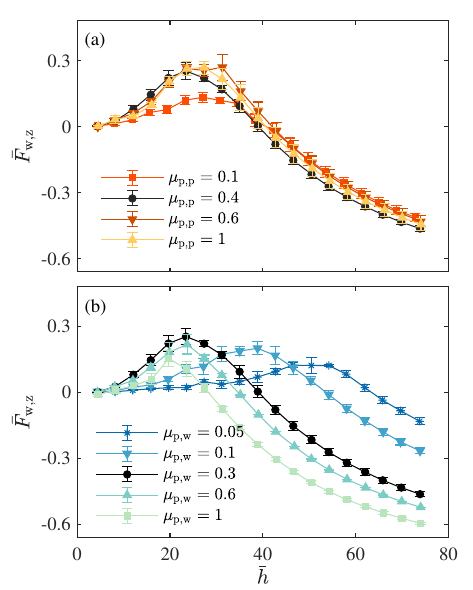}}

\caption{Wall force in the simulations where (a) particle-particle friction coefficient is modified, while particle-wall friction coefficient is kept fixed at its reference value, and (b) particle-wall friction coefficient is modified, while particle-particle friction coefficient is kept fixed at its reference value.
}
\label{fig:friction}
\end{figure}

Figure~\ref{fig:friction} shows separately the influence of particle-particle and particle-wall friction on the results. For the sake of simplicity, we focus here on the reference case where particles are released from a fixed height relative to the base. The results for other pouring heights are consistent and are not shown for conciseness. Figure~\ref{fig:friction}(a) shows the influence of particle-particle friction coefficient, keeping particle-wall friction fixed at its reference value. The dependence of $\bar F_{\rm w,z}^{\rm m}$ is monotonic—i.e., a larger friction coefficient leads to a stronger reverse Janssen effect without altering the location at which $\bar F_{\rm w,z}^{\rm m}$ is attained.  This behavior could be explained based on the fact that interparticle friction plays an important role in the momentum transfer from the impacting particles to the particles interacting with the side walls.  

Figure~\ref{fig:friction}(b) shows the influence of particle-wall friction.  Here we find more complicated behavior.  First, the dependence of  $\bar F_{\rm w,z}^{\rm m}$ is not monotonic any more; second, the peak of $\bar F_{\rm w,z}$ appears at a different location.  Non-monotonicity has also been found and discussed in~\cite{ciamarra_prl20}; however, the change of location of the maximum was not discussed (it should be noted that in~\cite{ciamarra_prl20}, apparently, both particle-particle and particle-wall friction coefficients were changed at once).  Figure~\ref{fig:friction} shows that the influence of the two quantities is fundamentally different.

\begin{figure}[!ht]
\centering \resizebox{1\hsize}{!}{\includegraphics{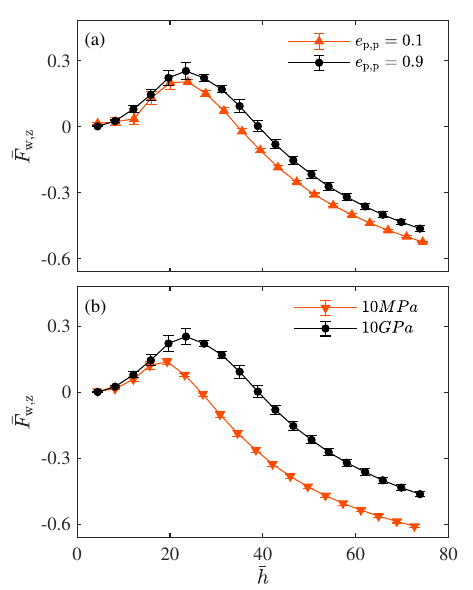}}
\caption{Wall force in the simulations when (a) restitution coefficient, and 
(b) Young's modulus of the particles is modified. 
}
\label{fig:particle}
\end{figure}

Figure~\ref{fig:particle}(a) shows the influence of the coefficient of restitution. Smaller values lead to a weaker reverse Janssen effect, as expected. 
Then, Fig.~\ref{fig:particle}(b) shows the influence of the particle Young's modulus.  For softer particles, there is less momentum transfer in the radial direction,  and therefore reverse Janssen effect weakens.  

\subsubsection{Influence of particle size and column diameter}

In this section,  we discuss the influence of particle size and column diameter;  as we will see, this discussion confirms further our understanding of the source of the reverse Janssen effect.  

When changing either particle size or column diameter, it is essential to establish the protocol and determine which quantities will be kept constant.  In order to remain consistent with the experiment, we choose to keep the total mass of the particles in each filling constant; therefore,  the number of particles delivered,  as well as (nondimensional) filling height, vary depending on particle size and column diameter.

\begin{figure}[!ht]
\centering \resizebox{1\hsize}{!}{\includegraphics{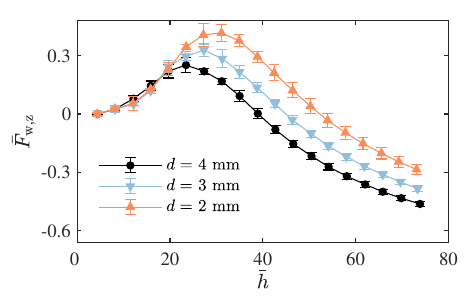}}
\caption{Wall forces exerted by the particles as a function of filling height for different particle sizes. Here, the ratio of tube diameter to particle size is kept fixed as $\bar D = 13$.
} 
\label{fig:fix_D_ver_d}
\end{figure}

Figure~\ref{fig:fix_D_ver_d} shows the results where the particle size is changed, but the ratio of tube diameter to particle size is kept fixed.  One may expect the collapse of the results (since it is typically considered that the ratio of the tube to particle size is a relevant parameter), but this is not the case: smaller particles lead to larger $\bar F_{\rm w,z}^{\rm m}$, which is furthermore shifted towards larger filling heights.  Recalling our understanding of the origin of the reverse Janssen effect, we could rationalize this trend of the results as follows.  First, the total energy delivered by the falling particles in each filling is essentially the same (independent of the particle size). However, if already deposited particles are smaller in size, it is easier to agitate them and move them towards the side walls; this is the reason for increased $\bar F_{\rm w,z}^{\rm m}$ as particle size decreases.  Regarding the filling height at which $\bar F_{\rm w,z}^{\rm m}$ occurs, we recall that the positive wall forces (in the ${\rm z}$ direction) occur at some non-vanishing distance from the free surface since the compression force due to particle weight is needed to allow for the development of particle-wall friction forces.  Then, for smaller particles, the required compression force is reached deeper in the sample when the distance from the free surface is measured in particle diameters.  This is the reason for the shift of the maximum of  $\bar F_{\rm w,z}$ curve towards larger values of $\bar h$ for smaller particles.

\begin{figure}[!ht]
\centering \resizebox{1\hsize}{!}{\includegraphics{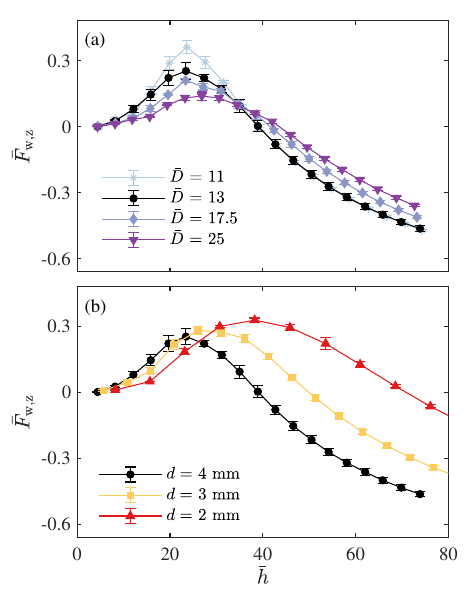}}
\caption{Wall force exerted by the particles versus the filling height for (a) particle size is kept fixed, and tube diameter is varied, and (b) particle size, $d$, is varied while the tube size is kept fixed at D = 52 mm.
}
\label{fig:D_ver_d}
\end{figure}

Figure~\ref{fig:D_ver_d}(a) shows the results where particle size is kept fixed, but (dimensional) tube radius increases.  As expected, we find a weaker reverse Janssen effect in this case, with $\bar F_{\rm w,z}^{\rm m}$ decreasing as tube diameter increases.  The explanation is that for larger tubes, it is more difficult for the agitated particles to reach the side walls.   

Figure~\ref{fig:D_ver_d}(b) shows perhaps more insightful case, where the dimensional tube diameter is kept fixed, but particle size is changed.  Here, as $d$ decreases, the ratio $D/d$ increases.  At first sight, one could expect a decrease of $\bar F_{\rm w,z}^{\rm m}$ for smaller particles.  However, it is important to note that there are two competing effects: in addition to the fact that the ratio of the column to particle size increases, smaller particles are easier to agitate, as illustrated by Fig.~\ref{fig:fix_D_ver_d}.  Since these two competing effects scale differently with the particle size (linear scaling of the ratio of particle size to tube diameter, and cubic scaling involving the particle inertia), the results in principle may depend on the exact values of the parameters considered.  For the present case, the effect related to particle inertia wins and leads to (slightly) larger $\bar F_{\rm w,z}^{\rm m}$ as particle size decreases.  Furthermore, the value of $\bar h$ where $\bar F_{\rm w,z}^{\rm m}$ is reached shifts to larger values for smaller values of $d$; careful analysis of the force distribution results (results analogous to the ones shown in Fig.~\ref{fig:wall_force}, additional figure not shown for brevity) show that this is due to the positive values of the wall force being larger and also appearing at the larger distances from the free surfaces for smaller particles.  Interestingly, when plotted in dimensional form, the results shown in this figure almost collapse; further research beyond the scope of this paper will be needed to understand such a collapse fully.

\subsection{Discussion of the source of difference between experimental and simulation results}
\label{sec:discussion}

One question that remains unanswered concerns the differences
between experimental and simulation results.  While the trends of some results are consistent (e.g., for wider tubes, keeping particle size the same, the reverse Janssen effect decreases in both experiments and simulations), there is a significant difference in the magnitude of the reverse Janssen effect for nominally the same conditions, compare, e.g., Figs.~\ref{fig:experiment}(c) and \ref{fig:simulations}(b) (We note here that the pouring height is larger in the experiments, however, this does not help understand the difference between experiments and simulations since larger pouring height suggests larger magnitude of the reverse Janssen effect, contrary to what is observed). Furthermore, we observe inconsistency between the experiments and simulation as particle-wall friction is modified; compare Figs.~\ref{fig:exp_force_varFrictionp} and~\ref{fig:friction}(b).  The details of the comparison are, however, complicated by a nonmonotonous dependence of the magnitude of the reverse Janssen effect on the particle-wall friction observed in simulations, as seen in Fig.~\ref{fig:friction}(b).  

Our interpretation is that the differences between experimental and simulation results is a combination of multiple factors, possibly involving the quantities that are known to influence the results but are difficult to control in experiments, such as the radial component of particle velocity and possibly the particle flux values.  In addition, there is an effect that we have not discussed so far:  in experiments, the energy is dissipated through multi-body interactions of the falling particles with multiple particles that are in the tube already, while simulations consider only interactions between particle pairs.  Stronger energy dissipation in experiments may lead to weaker momentum transport and, therefore, a weaker reverse Janssen effect.  Clearly, more work is needed to clarify the details of the observed differences between experiments and simulations.

\section{Conclusions}
\label{sec:concluston}

In this paper, we consider the origins of the reverse Janssen effect. We find that pouring particles from larger heights leads to a larger net upward force on the tube walls, leading to a base force that is larger than the weight of the material. While the details of the results may depend on the variety of particle and tube parameters, this counterintuitive finding is expected to hold as long as the tube diameter is relatively small when measured in units of particle diameters.  

The findings outlined in the preceding paragraph emphasize the importance of the kinetic energy of the poured particles. By reducing energy (most obviously by reducing the pouring height), one finds an order-of-magnitude weaker reverse Janssen effect, as confirmed by our simulations in which the pouring height was varied.  Furthermore, the simulations also illustrate the importance of collective effects, finding stronger reverse Janssen effect when the pouring particle flux is larger (while keeping the number of poured particles fixed).  
Interestingly, the experimental values (involving pouring from a significant height) show the reverse Janssen effect that is in between the simulation results obtained for small and large pouring heights. This finding suggests the need for more elaborate simulation models that would more accurately model the interaction of falling particles with granular beds and its consequences. We hope that our work will inspire such research directions.

\section*{Acknowledgement}
LK acknowledges useful discussion with Luis Pugnaloni. SS and AMMG acknowledge the financial support from the Research Council of Finland under grant No. 311138.

\end{document}